\documentclass[pra,twocolumn,showpacs,superscriptaddress]{revtex4}

\usepackage{amssymb}
\usepackage{amsmath}
\usepackage{amssymb}
\usepackage{bm}
\usepackage{lineno}
\usepackage{graphicx}

\newcommand{\upi}{\mathrm{i}}
\newcommand{\upe}{\mathrm{e}}

\newcommand{\sx}{\bm \sigma_{x}}

\newcommand{\sz}{\bm \sigma_{z}}

\newcommand{\be}{\begin{equation}}
\newcommand{\ee}{\end{equation}}
\newcommand{\bea}{\begin{eqnarray}}
\newcommand{\eea}{\end{eqnarray}}
\newcommand{\nn}{\nonumber}
\newcommand{\bfsfS}{\mbox{\sffamily{S}}}

\begin{document}

\title{Entanglement creation in circuit QED via Landau-Zener sweeps}

\author{Martijn Wubs}
\email{martijn.wubs@physik.uni-augsburg.de}
\author{Sigmund Kohler}
\author{Peter H{\" a}nggi}

\affiliation{Institut f\"{u}r Physik, Universit\"{a}t Augsburg,
D-86135 Augsburg, Germany}

\begin{abstract}
A qubit may undergo Landau-Zener transitions due to its coupling to
one or several quantum harmonic oscillators.  We show that for a
qubit coupled to one oscillator, Landau-Zener transitions can be
used for single-photon generation and for the controllable creation
of qubit-oscillator entanglement, with state-of-the-art circuit QED
as a promising realization. Moreover, for a qubit coupled to two
cavities, we show that Landau-Zener sweeps of the qubit are well
suited for the robust creation of entangled cavity states,  in
particular symmetric Bell states, with the qubit acting as the
entanglement mediator. At the heart of our proposals lies the
calculation of the exact Landau-Zener transition probability for the
qubit, by summing all orders of the corresponding series in
time-dependent perturbation theory. This transition probability
emerges to be independent of the oscillator frequencies, both inside
and outside the regime where a rotating-wave approximation is valid.
\end{abstract}


\maketitle

\section{Introduction}\label{sec:intro}
Entanglement is a purely quantum mechanical property of multipartite
systems. A system is entangled if its quantum state cannot be
described as  a direct product of states of its subsystems.
Entanglement is measurable in terms of nonclassical correlations of
the subsystems. Famous examples are the Einstein-Podolsky-Rosen
correlations between positions and momenta of two particles
\cite{Einstein1935a}, and the violation of the Bell inequalities by
spin systems that are described by Bell states \cite{Bell1964a}.
Fundamental tests of entanglement and nonlocality were performed in
quantum optics \cite{Aspect1981a}, with the  intriguing outcome that
measured nonclassical correlations between entangled spatially
separated subsystems rule out local realism \cite{Redhead1987a}.

With the advent of quantum information theory in recent years, the
interest in entanglement has broadened. Many efforts exist to make
use of entanglement in  information processing or to
quantum-communicate with built-in security \cite{Nielsen2000a}. With
quantum information processing in mind, in this paper we propose to
create entanglement in
 two spatially separated elements
(circuit cavities) of a superconducting circuit by letting a third
element (the superconducting qubit) undergo a Landau-Zener sweep.
This will be a robust method to create Bell states in two-cavity
circuit cavity quantum electrodynamics (QED).

In optical cavity QED,  atoms (qubits or $N$-level systems) become
entangled with optical cavity modes (quantum harmonic oscillators).
The creation of atom-cavity entanglement in cavity QED is possible
because the strong coupling regime can be realized, where Rabi
oscillations between atomic and optical excitations occur on a
faster time scale than spontaneous emission and cavity decay. By
adiabatic passage the quantum state of an atom flying through an
optical cavity can be mapped onto the quantum state of the cavity.
It was investigated theoretically \cite{Lange2000a,Ikram2002a} and
shown experimentally \cite{Rauschenbeutel2001a} that in a similar
manner also two modes of the same cavity can be entangled.
Alternatively, two spatially separated optical cavities could be
entangled by letting an atom fly successively through both cavities
\cite{Messina2002a,Browne2003a}, or by detecting  photons leaving
the two cavities \cite{Browne2003b}.

If the aim is to build devices for quantum information processing,
then scalability to several qubits and oscillators is an important
prerequisite. It seems technologically very challenging to scale up
optical cavity QED. Recently, a new research field called circuit
QED has emerged in which analogues of cavity QED have been realized
with superconducting qubits \cite{Nakamura1999a,VanderWal2000a} and
oscillators. A flux qubit was  coupled strongly to a superconducting
quantum interference device \cite{Chiorescu2004a}, and a charge
qubit to a transmission line resonator
\cite{Wallraff2004a,Blais2004a}, and Rabi oscillations have been
observed in both experiments. There are analogous proposals to
couple  superconducting qubits to nanomechanical resonators
\cite{Cleland2004a,Geller2005a}. Superconducting circuits are
promising  because of their potential scalability and because many
of their parameters are tunable over a broad range.

Scaling up present-day circuit QED can be done by increasing the
number of qubits, the number of oscillators, or both. The first
option is  encountered in most proposals, where the quantum harmonic
oscillator is to be used as a bus to manipulate and read out the
qubits \cite{Cleland2004a,Geller2005a,Migliore2006a,Deng2006a}. The
second option is the circuit analogue of two-cavity (or $N$-cavity)
QED \cite{Messina2002a,Browne2003a}, and the present paper belongs
to this category. Also in this group is the recent proposal to
couple a superconducting qubit to two transmission-line resonators
with orthogonal field polarizations \cite{Storcz2006a}.
 As to the third option, examples of proposals for experiments with several
superconducting qubits coupled to several oscillators can be found
in \cite{Geller2005a,Zhang2006a}. Probably the first realization of
such a complex setup will consist of two qubits independently
coupled to their own oscillators, allowing independent qubit state
manipulation and readout.

One method to  manipulate the state of an isolated qubit is to use
Landau-Zener sweeps \cite{Landau1932a,Zener1932a,Stueckelberg1932a},
as discussed in more detail below.   LZ transitions can be used to
control qubit gate operations \cite{Saito2004a,Hicke2006a} and to
read out qubits \cite{Ankerhold2003a}.  Recently, LZ transitions
have been observed in various experiments with superconducting
qubits \cite{Izmalkov2004a,Ithier2005a,Oliver2005a,Sillanpaa2006b}.

In this paper we concentrate on quantum state manipulation in
multi-cavity circuit QED  via Landau-Zener sweeps of a qubit.   Bit
flips in the qubit  can take place even in the absence of a direct
coupling between the qubit levels, induced instead by the coupling
to the oscillators. Oscillators in highly excited coherent states
can be described classically and give rise to a non-monotonic LZ
transition probability as a function of the coupling
strength~\cite{Wubs2005a}. Here we focus on the experimentally
relevant  situation in which the oscillators all start in their
ground states. We show how single photons can be created, not only
in the presence of only one oscillator as in our prior
work~\cite{Saito2006a}, and  how symmetric Bell states can be
created in two circuit oscillators.

The decisive advantage of our proposal is that qubit-oscillator
interaction strengths are
 static, in contrast to the situation in standard cavity QED
 where these interactions require
precise dynamical control \cite{Messina2002a,Ikram2002a,Deng2006a}.
Other advantages of LZ transitions are that even adiabatic
interactions can be performed rather fast \cite{Ashhab2006a}, and
that  LZ transition probabilities are extremely robust under
dephasing \cite{Ao1989a,Wubs2006a}.

In Sec.~\ref{standard} we
 review the standard LZ problem, before deriving in
Sec.~\ref{Sec:LZNosc} exact LZ transition probabilities for a qubit
coupled to $N$ harmonic oscillators. This result is used in
Sec.~\ref{sec:singleosc} about single-photon generation and in the
central section~\ref{sec:twoosc} about the generation of  entangled
cavity states via LZ sweeps of the qubit. The
realization in circuit QED of our proposals  is discussed
in Sec.~\ref{sec:exp}, before we conclude.

\section{Standard Landau-Zener transition}\label{standard}
For later use,  we shall briefly review the well-known Landau-Zener
transition in a qubit. Consider a two-state system with states
$|{\uparrow}\rangle$ and $|{\downarrow}\rangle$. The standard
Landau-Zener(-Stueckelberg) problem is defined by the Hamiltonian
\cite{Landau1932a,Zener1932a,Stueckelberg1932a}
\begin{equation}\label{H_LZstandard}
H(t) = \frac{v t}{2}\sz +  \Delta\sx,
\end{equation}
where $\sz \equiv |{\uparrow}\rangle\langle {\uparrow}| -
|{\downarrow}\rangle\langle {\downarrow}|$ and $\sx \equiv
|{\uparrow}\rangle\langle {\downarrow}| +|{\downarrow}\rangle\langle
{\uparrow}|$ are Pauli matrices. This Hamiltonian describes how at
time $t=0$ the diabatic energies $\pm vt/2$ of the two states cross
with level-crossing speed $v$. During the crossing, the two diabatic
states interact with a strength $\Delta$, so that the adiabatic
states (or: time-dependent eigenstates) differ from the diabatic
states. As is usual, the  adiabatic energies are found as the
time-dependent eigenvalues of the Hamiltonian~(\ref{H_LZstandard}).
These are  $\pm\sqrt{(v t/2)^{2}+\Delta^{2}}$, showing the
archetypical avoided level crossing. The gap between these two
adiabatic energies  is at least $2\Delta$, and the minimum occurs at
time $t=0$.  More intriguing is the dynamics of the state
\begin{equation}\label{TLstate}
|{\psi(t)}\rangle = c_{\uparrow}(t)|{\uparrow}\rangle
+c_{\downarrow}(t)|{\downarrow}\rangle = \left(\begin{array}{c} c_{\uparrow}(t)\\
c_{\downarrow}(t) \end{array}\right),
\end{equation}
as described by the Hamiltonian~(\ref{H_LZstandard}). Except around
$t=0$,  the Hamiltonian (\ref{H_LZstandard}) is dominated by the
diabatic energies. It  therefore makes sense to define an
interaction picture by the transformation $U_{0}(t) = \exp(-\upi
vt^2\sz/2\hbar)$, that is $|\psi(t)\rangle =
U_{0}(t)|\tilde{\psi}(t)\rangle$ and $|\tilde{\psi}(t)\rangle=[
\tilde{c}_1(t)|1\rangle + \tilde{c}_2(t)|2\rangle]$, where the
interaction-picture probability amplitudes obey
\begin{equation}
\label{eqmotint} \frac{\mbox{d}}{\mbox{d}t}
\begin{pmatrix}\tilde{c}_{\uparrow} \\ \tilde{c}_{\downarrow} \end{pmatrix}
= -\frac{\mathrm{i}}{\hbar}
  \begin{pmatrix} 0 & \Delta\;\upe^{\mathrm{i} v t^{2}/2\hbar}  \\
  \Delta\;e^{-\mathrm{i} v t^{2}/2\hbar} & 0 \end{pmatrix}
  \left(\begin{array}{c}\tilde{c}_{\uparrow} \\ \tilde{c}_{\downarrow} \end{array} \right).
\end{equation}
This system of coupled equations of motion (\ref{eqmotint}) is one
of the few in driven quantum mechanics that can be solved exactly.
The dynamics at all times  can be expressed in terms of parabolic
cylinder functions \cite{Zener1932a}. Instead of presenting this
dynamics in full glory here, we will give the following useful
summary of the dynamics
\begin{equation}
|{\tilde\psi(t=\infty)}\rangle =
\bfsfS_{\Delta}|{\psi(t=-\infty)}\rangle,
\end{equation}
in terms of the exact scattering matrix (or S-matrix)
\begin{equation}\label{transitionSb}
\bfsfS_\Delta = \left( \begin{array}{cc} \sqrt{q} &
\sqrt{1-q}\;\upe^{-\mathrm{i}\chi}
\\- \sqrt{1-q}\;\upe^{\mathrm{i} \chi} & \sqrt{q} \end{array}\right).
\end{equation}
Here the quantity $q$ stands for $\exp(-2\pi \eta)$ and the Stokes
phase $\chi = \pi/4+\arg\Gamma(1-\mathrm{i}\eta)+\eta(\ln \eta-1)$,
with adiabaticity parameter $\eta = \Delta^{2}/(\hbar v)$ and
$\Gamma$ the Euler Gamma function.
It follows that the probability $P_{\downarrow}$ that the atom ends
up in the initially unoccupied level $|{\downarrow}\rangle $ is
given by
\begin{equation}
 \label{standardpLZ}
 P_{\downarrow}\equiv  |c_{\downarrow}(t=\infty)|^2 = 1 - \upe^{-2\pi \Delta^{2}/\hbar v}.
\end{equation}
This is the famous Landau-Zener transition probability. It is an
exact result for all  $\Delta$ and $v$. Instead of using the
properties of parabolic cylinder functions to derive this result,
which is the standard method, the same transition probability for
the standard Landau-Zener problem can be found by exact summation of
an infinite series in time-dependent perturbation theory
\cite{Kayanuma1984a}. The latter method is less well known, but it
is an important one, since it is this perturbation method that can
be used to analyze Landau-Zener transitions  in more complex systems
as well \cite{Volkov2005a,Dobrescu2006a,Wubs2006a}, where an
analysis in terms of special functions is not available. An example
of a more complex system analyzed by a perturbation series is
presented below.

\section{Landau-Zener transitions of a qubit coupled to many
oscillators}\label{Sec:LZNosc}

Next we turn to our main topic, namely LZ sweeps in a qubit that is
coupled to $N$ cavity modes. We assume that the internal interaction
$\Delta$ of the qubit vanishes, unlike in the standard LZ
Hamiltonian of Eq.~(\ref{H_LZstandard}), so that all bit flips that
occur in the qubit are mediated by the cavity modes. The latter are
described as quantum harmonic oscillators. The Hamiltonian becomes
\begin{equation}\label{HtNosc}
H = \frac{vt}{2} \sigma_{z}
   + \sigma_{x}\sum_{j=1}^N \gamma_j (b_{j} + b_{j}^{\dagger})
   + \sum_{j=1}^N \hbar\Omega_{j} b_{j}^{\dagger} b_{j} \; .
\end{equation}
By keeping $N$ general here, our calculations are  relevant both for
setups with one and with two oscillators, on which we focus in later
sections. Moreover, in Sec.~\ref{sec:exp} we will show that this
Hamiltonian~(\ref{HtNosc}) indeed describes the dynamics for LZ
sweeps in circuit QED.

We will now calculate the LZ transition probability for a qubit that
at time $t=-\infty$ starts in its ground state $|{\uparrow}\rangle$.
We assume that all cavity modes are initially in their ground states
$|\bm 0\rangle$ as well, where $|\bm n\rangle$ is a shorthand
notation for the Fock state $|n_1,\ldots,n_N\rangle$.
Since we chose $\Delta=0$ so that the states $|{\uparrow}\rangle$
and $|{\downarrow}\rangle$ are eigenstates of the qubit Hamiltonian
$\frac{1}{2}vt\sigma_z$.  Then any transition between the two qubits
states can only result from the coupling to the oscillator. Notably,
our model \eqref{HtNosc} is significantly different from the
``standard'' Landau-Zener problem extended by a bath coupling via
$\sigma_z$ \cite{Ao1989a, Kayanuma1998a}.

The central quantity of interest is the probability
$P_{\uparrow\to\downarrow}(t)$ that the qubit has flipped to the
state $|{\downarrow}\rangle$.  If the qubit energy is switched
slowly (i.e. $v\to 0$), the qubit will follow the adiabatic ground
state which at large times is the state $|{\downarrow}\rangle$; then
$P_{\uparrow\to\downarrow}(\infty) = 1$.  For large $v$, the qubit
will remain in state $|{\uparrow}\rangle$, so that
$P_{\uparrow\to\uparrow}(\infty) = 1$ and
$P_{\uparrow\to\downarrow}(\infty) = 0$, corresponding to a
nonadiabatic transition. Generalizing the calculation presented in
our prior work~\cite{Saito2006a} from one to arbitrarily many
oscillators, we derive in the following an exact expression for
$P_{\uparrow\to\uparrow}(\infty) = \sum_{\bm n}
|\langle{\uparrow},\bm n| U(\infty,-\infty) |{\uparrow},\bm 0\rangle
|^2$ where $U(t,t')$ denotes the time-evolution operator.

We start by a transformation to an interaction picture with respect
to the uncoupled qubit and oscillators,
\begin{equation}
U_0(t) = \exp\Big(-\upi\sum_j\Omega_j b_{j}^\dagger b_{j} t\Big)
\exp\Big(-\frac{\upi}{4\hbar}vt^2 \sigma_z\Big)
\end{equation}
which yields the interaction-picture Hamiltonian
\begin{equation}
\label{Hint} \tilde H(t) = \sum_j\gamma_j
(b_j^\dagger\upe^{\upi\Omega_j t} + b_j\upe^{-\upi\Omega_j t})
  \exp\Big(-\frac{\upi}{2\hbar}vt^2\sigma_z\Big) \sigma_x .
\end{equation}
Within a perturbation expansion of the probability amplitude $A_{\bm
n} = \langle{\uparrow},\bm n| U(\infty,-\infty) |{\uparrow},\bm
0\rangle$, we obtain the series $A_{\bm n} = \sum_{k=0}^\infty
(-\upi/\hbar)^{2k} a_{\bm nk}$. Since each $\tilde H(t)$ flips the
qubit exactly once, only even orders of $\gamma$ appear in $A_n$.
The $2k$-th order contribution $a_{\bm{n}k}$ is characterized by
$2k$ vectors $\bm{\lambda}_\ell$, $\ell=1,\ldots,2k$, where
$\bm\lambda$ denotes a vector with exactly one component equal to
$\pm 1$, while all other components vanish.  The $j$th component
$(\bm\lambda_\ell)_j=\pm 1$ corresponds to the operators
$b_j^\dagger$ and $b_j$, respectively. Then we obtain
\begin{equation}
\label{ank}
\begin{split}
a_{\bm nk} =& \sum_{\bm\lambda_{2k}\cdots\bm\lambda_1} \!\!\!\!
C_{nk}(\lambda_{2k},...,\lambda_1) \! \int\limits_{-\infty}^\infty
\!\mbox{d}t_{2k}\! \int\limits_{-\infty}^{t_{2k}} \!\mbox{d}t_{2k-1}
\cdots \!\! \int\limits_{-\infty}^{t_2} \!\mbox{d}t_1
\\
&\times
\exp\Big[\upi \sum_{\ell=1}^{2k} \bm\Omega\cdot \bm\lambda_\ell t_\ell
         +\frac{\upi v}{2\hbar} \sum_{\ell=1}^k (t_{2\ell}^2 - t_{2\ell-1}^2)
    \Big] ,
\end{split}
\end{equation}
where we introduced $\bm\Omega=(\Omega_1,\ldots,\Omega_N)$ in order
to obtain a more compact vector notation.  The appearence of the
$\bm\lambda_\ell$ in the exponent stems from the sign in the
time-dependent phase of the creation and annihilation operators. The
dots in the coefficient $C_{\bm nk}(\lambda_{2k},\ldots,\lambda_1) =
\langle \bm n|\cdots|\bm 0\rangle$ denote the combination of $2k$
operators $\gamma_j b_j$ and $\gamma_j b_j^\dagger$ that corresponds
to the sequence $\bm\lambda_{2k},\ldots,\bm\lambda_1$.
An important simplification of the $\bm\lambda$-summation results from
the fact that $C_{\bm nk}=0$ whenever more annihilation than
creation operators act on the oscillator ground state $|\bm 0\rangle$.
Thus, we need to consider only those $\bm\lambda$-sequences that for all
$\ell \leq 2k$ fulfill the relation
\begin{equation}
\sum_{\ell'=1}^{\ell} \sum_j (\bm\lambda_{\ell'})_j \geq 0.
\label{cond1}
\end{equation}

\begin{figure}
\centerline{\includegraphics{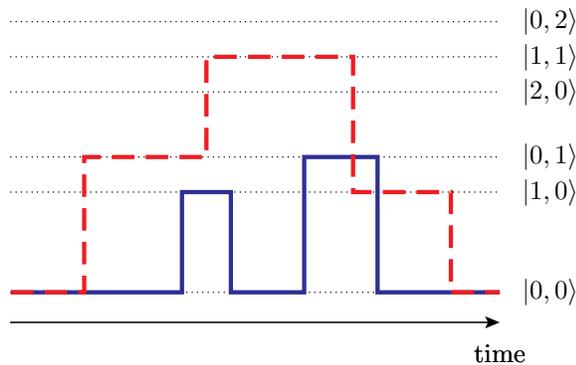}}\label{fig:perturbation}\caption{(Color
online) Possible processes during LZ transitions in two-cavity
circuit QED. Shown are the lowest-energy levels of the harmonic
oscillator states $|n_1,n_2\rangle$ for $N=2$. The solid blue line
marks a process that contributes to the perturbation series for
Landau-Zener transition probability
$P_{{\uparrow}\to{\uparrow}}(\infty)$: the oscillator state jumps
(repeatedly) from the ground state to an arbitrary one-photon state
and then back to the ground state, in agreement with the no-go-up
theorem. In contrast, the process marked by the red dashed line
contributes to $P_{\uparrow\rightarrow\uparrow}(t)$ only at finite
times.}
\end{figure}

For the further evaluation, we substitute in Eq.~\eqref{ank} the times
$t_\ell$ by the time differences $\tau_\ell = t_{\ell+1}-t_\ell$,
$\ell = 1,\ldots,2k-1$, where $t=t_{2k}$.  Thus, we insert $t_\ell = t -
\sum_{\ell'=\ell}^{2k-1} \tau_{\ell'}$ so that the integral in
\eqref{ank} becomes
\begin{equation}
\begin{split}
\label{int} \int\limits_{-\infty}^\infty \mbox{d}t
\int\limits_0^\infty \mbox{d}\tau_{2k-1}\ldots \mbox{d}\tau_1
\exp\Big[\upi\sum_{\ell=1}^{2k} \bm\Omega\cdot\bm\lambda_\ell
\Big(t-\sum_{\ell'=\ell}^{2k-1} \tau_{\ell'} \Big)\Big]
\\
\times
\exp\Big[\frac{\upi v}{2\hbar} \sum_{\ell=1}^k \Big\{ 2\tau_{2\ell-1}
\Big(t - \sum_{\ell'=2\ell}^{2k-1} \tau_{\ell'} \Big) - \tau_{2\ell-1}^2
\Big\}
\Big] .
\end{split}
\end{equation}
The $t$-integration results in the delta function
\begin{equation}\label{deltafunction}
2\pi\, \delta\Big( \frac{v}{\hbar}\sum_{\ell=1}^k \tau_{2\ell-1}
+ \sum_{\ell=1}^{2k} \bm\Omega\cdot\bm\lambda_\ell\Big) .
\end{equation}
From the inequality \eqref{cond1} it follows that the second sum in
the argument of the delta function is non-negative.  Because the
integration interval of all $\tau_\ell$ is $[0\ldots\infty)$, any
non-zero contribution to the integral \eqref{int} results from
$\tau_1 = \tau_3 = \ldots = \tau_{2k-1} = 0$.  Hence, the integral
over the time differences $\tau_2, \tau_4, \ldots, \tau_{2k-2}$ must
yield a distribution proportional to
$\delta(\tau_1)\,\delta(\tau_3)\cdots \delta(\tau_{2k-1})$.
Evaluating the integrals over all $\tau_{2\ell}$ separately, one
finds that such a distribution is obtained only if
$\sum_{\ell'=1}^{2\ell}(\bm\lambda_{\ell'})_j = 0$ for all
$\ell=1,\ldots, k-1$.  These $k-1$ relations together with the delta
function~\eqref{deltafunction} lead to the conditions
$(\bm\lambda_{2\ell}+\bm\lambda_{2\ell-1})_j = 0$ and hence
$\bm\lambda_{2\ell-1}=-\bm\lambda_{2\ell}$ for all
$\ell=1,\ldots,k$.  In combination with Eq.~\eqref{cond1} this
implies that an integral is non-vanishing only if the non-zero
component of $\bm\lambda_{2\ell-1}$ is $+1$ while the same component
of $\bm\lambda_{2\ell}$ equals $-1$. In other words, we obtain the
selection rule that to the occupation probability at $t=\infty$ only
those processes contribute in which the oscillator jumps
(repeatedly) from the state $|\bm 0\rangle$ to any state with a
single photon (i.e. to $b_{j}^{\dag}|\bm 0\rangle$) and back; see
Fig.~\ref{fig:perturbation}. It follows that the oscillators not
only start but also end in their ground state $|\bm 0\rangle$ if the
final qubit  state is $|{\uparrow}\rangle$. We call this dynamical
selection rule the ``no-go-up theorem'' (see also
\cite{Volkov2005a}).

Stating the above result in mathematical terms,  of all possible
$2k$-th order processes denoted by sequences of vectors
$-\bm\lambda_{2k-1} ,\bm\lambda_{2k-1}, \ldots, -\bm\lambda_3,
\bm\lambda_3, -\bm\lambda_1, \bm\lambda_1$, the processes that
contribute to the perturbation series for
$P_{\uparrow\to\uparrow}(\infty)$ can be characterized by simpler
sequences of scalars $j_k, \ldots, j_2, j_1$, where $j_\ell$ is the
index of the non-vanishing component of both $\bm\lambda_{2\ell-1}$
and $\bm\lambda_{2\ell}$. All prefactors $C_{\bm nk}$ of
contributing processes have the structure
\begin{align}
C_{\bm nk}
&= \gamma_{j_k}^2\ldots\gamma_{j_3}^2\gamma_{j_1}^2
    \langle \bm n|b_{j_{2k-1}}b_{j_{2k-1}}^\dagger\ldots
    b_{j_1}b_{j_1}^\dagger|\bm 0\rangle
\nonumber \\
&= \gamma_{j_{2k-1}}^2 \ldots\gamma_{j_3}^2 \gamma_{j_1}^2 \delta_{\bm n,\bm 0}.
\end{align}

The remaining multiple integrations are performed as detailed in the
appendix of Ref.~\cite{Kayanuma1984a}, yielding  $a_{\bm nk} =
\delta_{\bm n,0} (\pi\hbar/v)^k/k!$ and is independent of the indices
$j_\ell$.  Therefore the summation over $j_1,\ldots,j_k$ can be
identified as $(\sum_{j=1}^N \gamma_j^2)^k$ so that
\begin{equation}
A_{\bm n} = \delta_{\bm n,\bm 0} \exp[-\pi S/\hbar v).
\end{equation}
Consequently, we  find the transition probability
\begin{equation}
\label{centralresult2} P_{\uparrow\to\downarrow}(\infty) =
1-P_{\uparrow\to\uparrow}(\infty) = 1-\upe^{-2\pi S/\hbar v} \;,
\end{equation}
in terms of the integrated spectral density
\begin{equation}\label{Sdef}
S \equiv \sum_{j = 1}^{N} \gamma_{j}^{2}
\end{equation}
The final transition probability~\eqref{centralresult2} depends only
on $S$ which acts as the effective coupling strength. Notice that
quite surprisingly, the transition probability
$P_{\uparrow\to\downarrow}(\infty)$ is independent of the oscillator
frequencies $\Omega_j$.  Nevertheless, the dynamics at intermediate
times does depend on them. This was shown to be the case for  a
qubit coupled to one oscillator in \cite{Saito2006a}, and it will
also hold for setups with two oscillators, as  presented in
Sec.~\ref{sec:twoosc}.

\section{Manipulation of the single-oscillator
state}\label{sec:singleosc}
We first consider LZ transitions in the standard cavity QED model
of one qubit coupled to one oscillator.
 Since we start out in the ground state
$|{\uparrow},0\rangle$ and the Hamiltonian in Eq.~\eqref{Hint}
correlates every creation or annihilation of a photon with a qubit
flip, the resulting dynamics is restricted to the states
$|{\uparrow},2n\rangle$ and $|{\downarrow},2n{+}1\rangle$.
As derived above, of the former states only $|{\uparrow},0\rangle$
stays occupied.
Thus, the final state exhibits
a peculiar type of entanglement between the qubit and the
oscillator, and can be written as
\begin{equation}
\begin{split}
|\Psi(\infty)\rangle
={}&
  \sqrt{1-P_{{\uparrow}\to{\downarrow}}(\infty)}\,|{\uparrow}0\rangle
\\ &
 +\sqrt{P_{{\uparrow}\to{\downarrow}}(\infty)}\,
  \big(c_1|{\downarrow}1\rangle + c_3|{\downarrow}3\rangle
       +\ldots\big) ,
\end{split}
\end{equation}
where $|c_1|^2+|c_3|^2+\ldots = 1$.  This implies that by measuring
the qubit in state $|{\downarrow}\rangle$, a highly nonclassical
oscillator state is produced in which only odd-photon (or:
odd-phonon) states are occupied.  Qubit and oscillator end up fully
entangled, in the sense that after tracing out the oscillator
states, no coherence between the qubit states $|{\uparrow}\rangle$
and $|{\downarrow}\rangle$ is left. The entanglement would have been
less perfect if a nonvanishing internal interaction $\propto\,\sx$
between the qubit states $|{\uparrow}\rangle$ and
$|{\downarrow}\rangle$ had been present \cite{Wubs2006a}.

\begin{figure}[t]
\begin{center}
\includegraphics*[width=8cm]{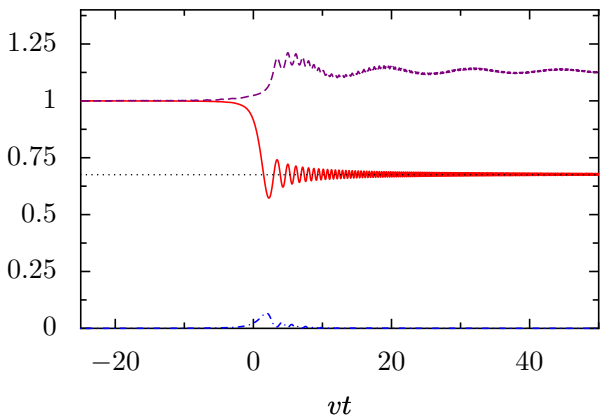} \end{center}
\caption{(Color online) LZ dynamics of a qubit coupled to one
oscillator, far outside the RWA regime: $\gamma = \hbar\Omega =
0.25\sqrt{\hbar v}$. The red solid curve is the survival probability
$P_{{\uparrow}\to{\uparrow}}(t)$ when starting in the initial state
$|{\uparrow 0}\rangle$. The dotted black line is the exact survival
probability $P_{\uparrow \rightarrow \uparrow }(\infty)$ based on
Eq.~(\ref{centralresult2}). The dashed purple  curve depicts the
average photon number in the oscillator if the qubit would be
measured in state $|{\downarrow}\rangle$; the dash-dotted blue curve
at the bottom shows the analogous average photon number in case the
qubit would be measured $|{\uparrow}\rangle$.
\label{fig:photon_averages}}
\end{figure}

While $P_{{\uparrow}\to{\downarrow}}(\infty)$ is determined by the
ratio $\gamma^2/\hbar v$, the coefficients $c_{2n+1}$ depend also on
the oscillator frequency. In Fig.~\ref{fig:photon_averages} we
depict how for a small frequency (very small: equal to the coupling
strength!) the average photon numbers in the oscillator depend on
the state of the qubit.
In particular,  the average photon number decays rapidly to zero in
case the qubit ends ``up'', in agreement with the no-go-up theorem
derived in the previous section, whereas on average more than one
photon resides in the oscillator in case the qubit has flipped to
$|{\downarrow}\rangle$. Furthermore, it can be seen that the
probability to end up in $|{\uparrow}\rangle$ indeed tends to the
analytically exact final value.

In contrast to the extreme situation depicted in
Fig.~\ref{fig:photon_averages}, for the recent experiments in
circuit QED \cite{Wallraff2004a} and for optical cavity QED one is
always in the situation $\gamma\ll\hbar\Omega$, in which case
$c_1\approx 1$ to a very good approximation. Hence one can control
via $v$ the final state to be any superposition of
$|{\uparrow}0\rangle$ and $|{\downarrow}1\rangle$. In particular, in
the adiabatic limit $v\hbar/\gamma^2 \ll 1$, the final state becomes
$|{\downarrow}1\rangle$. This has the important physical implication
of the creation of exactly one photon in the cavity, triggered by a
Landau-Zener transition.  In an experiment, the photon will
subsequently leak out of the cavity.

By exploiting these two processes, we propose the following
four-step LZ cycle for single-photon generation in circuit QED, as
sketched in Fig.~\ref{fig:single_photon_cycle}:
\begin{figure}[t]
\begin{center}
\includegraphics*[width=8cm]{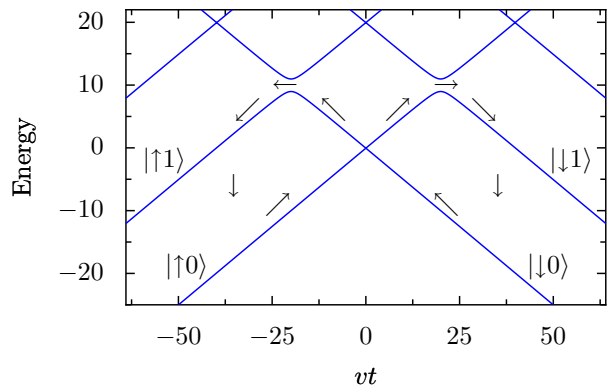} \end{center}
\caption{(Color online) Sketch of adiabatic eigenstates during LZ
sweep of a qubit that is coupled to
 one oscillator. Starting in the ground state $|{\uparrow 0}\rangle$ and
 by choosing a slow LZ sweep, a single photon can be created in the oscillator.
 Due to cavity decay, the 1-photon state will decay to a zero-photon state. Then the reverse LZ sweep
 creates another single photon that eventually decays to the initial state $|{\uparrow 0}\rangle$. This
 is a cycle to create single photons that can be repeated. } \label{fig:single_photon_cycle}
\end{figure}
The first step is single-photon generation in the cavity via the
adiabatic LZ transition $|{\uparrow}0\rangle \rightarrow
|{\downarrow}1\rangle$, brought about by switching the Josephson
energy sufficiently slowly. Second, the photon is released from the
cavity via the (controlled) cavity decay $|{\downarrow}1\rangle
\rightarrow |{\downarrow}0\rangle$. In the third step, another
individual photon is generated via the reverse LZ sweep
$|{\downarrow}0\rangle \rightarrow |{\uparrow}1\rangle$. Fourth and
finally, a further photon decay completes the cycle.

This scheme for repeated photon generation via Landau-Zener cycles
could be implemented in circuit QED, where the atom-cavity coupling
remains at a constant and high value and where qubits are highly
tunable so that LZ sweeps can be made from minus to plus an
``atomic'' frequency, and back. Details about the physical
realization in circuit QED of this proposal  are deferred to
Sec.~\ref{sec:exp}. The strength of this scheme is its simplicity
and its robustness against parameter variations, especially
variations (and fluctuations \cite{Wubs2006a}) of the oscillator
frequency.

\section{Entangling two cavities by a LZ sweep of a
qubit}\label{sec:twoosc}
 Now consider the situation that the qubit
is coupled to {\em two} cavities instead of one, with resonance
frequencies $\Omega_{1,2}$. We will now show that the two cavity
oscillators become entangled by a Landau-Zener sweep of the qubit,
and moreover that the specific entangled state in which the
oscillators end up can be engineered by varying the Landau-Zener
sweep speed $v$ and the frequency detuning
$\delta\omega=(\Omega_{2}-\Omega_{1})$ of the oscillators.

 As before in the case of one
oscillator, we will assume $\Delta=0$ for the internal interaction
of the qubit, so that all bit flips in the qubit are caused by
interactions with the oscillators. For simplicity, we will assume
that the two qubit-oscillator strengths are equal,
$\gamma_{1}=\gamma_{2}=\gamma$. Without loss of generality, we can
take $\Omega_{1} \le \Omega_{2}$ so that the detuning $\delta\omega$
is nonnegative. The Hamiltonian becomes
\begin{equation}\label{Ht2}
H = \frac{vt}{2} \sigma_{z}
   + \gamma\sigma_{x} (b_{1} + b_{1}^{\dagger}+b_{2} + b_{2}^{\dagger})
   + \hbar\Omega_{1} b_{1}^{\dagger} b_{1} + \hbar\Omega_{2} b_{2}^{\dagger} b_{2}.
\end{equation}
We will assume that this system starts in the ground state
$|{\uparrow 0_{1} 0_{2}}\rangle$ before undergoing the Landau-Zener
transition. (The subscripts $1,2$ that label the two oscillators
will be left out below.) The general result~(\ref{centralresult2})
then implies that the probability for the qubit to end up in the
state $|{\downarrow}\rangle$ equals
\begin{equation}
\label{centralresulttwoosc} P_{\uparrow\to\downarrow}(\infty) =1-
P_{\uparrow\to\uparrow}(\infty)= 1-\upe^{-2\pi
(\gamma^{2}+\gamma^{2})/\hbar v}.
\end{equation}
At this point it is important to appreciate that this exact result
has been derived without making a rotating-wave approximation and by
taking the full Hilbert space of the two oscillators into account.
The absence of any frequency dependence in
Eq.~(\ref{centralresulttwoosc})  is therefore quite surprising.

\begin{figure}[t]
\begin{center}
\includegraphics*[width=8cm]{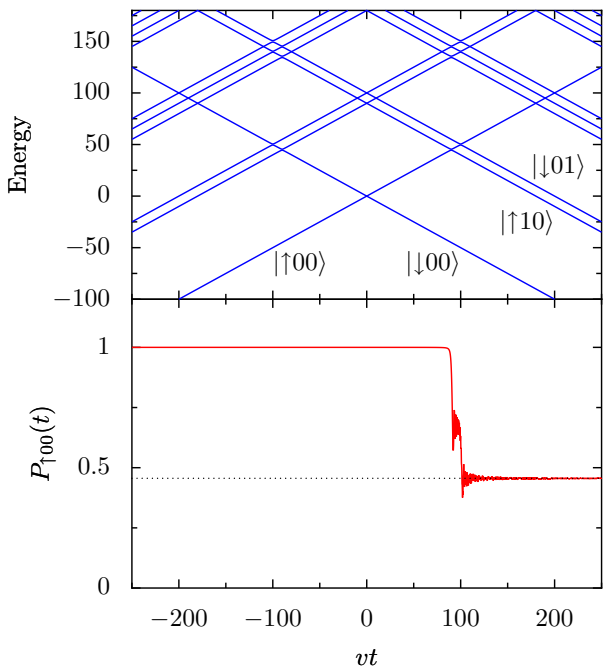} \end{center}
\caption{(Color online)  Upper panel: Adiabatic energies during a LZ
sweep of a qubit
 coupled to two oscillators. Parameters:
$\gamma = 0.25\sqrt{\hbar v}$, $\hbar\Omega_{1}= 90\sqrt{\hbar v}$
and $\Omega_{2}= 100\sqrt{\hbar v}$. Viewed on this scale of
oscillator energies, the differences between exact and avoided level
crossings are invisible. Lower panel: for the same parameters,
probability $P_{\uparrow\rightarrow\uparrow}(t)$ that the system
stays in the initial state $|{{\uparrow} 0 0}\rangle$ (solid), and
corresponding exact survival final survival probability
$P_{\uparrow\rightarrow\uparrow}(\infty)$ of
Eq.~(\ref{centralresulttwoosc}) (dotted).}
\label{fig:energylandscape}
\end{figure}

In the following we are interested in the properties of the final
qubit-two-oscillator state $|{\psi(\infty)}\rangle$ rather than
merely the transition probability
$P_{\uparrow\to\downarrow}(\infty)$ of the qubit. In general not
much can be said about this final state, but let us now make the
realistic assumption $\hbar\Omega_{1,2} \gg \gamma$: both oscillator
energies
 $\hbar\Omega_{1,2}$ are much larger than the qubit-oscillator
 couplings $\gamma_{1}=\gamma_{2}=\gamma$. Still, the frequency {\em
 detuning} $\delta\omega=(\Omega_{2}-\Omega_{1})$ may be larger or smaller than
$\gamma/\hbar$.  The adiabatic energies
 in this case are sketched in Fig.~\ref{fig:energylandscape}.
 As shown in the figure, level crossings
 that are important for the final state only occur around the times when the qubit energy
 $v t$ is resonant with one or both of the oscillator energies $\hbar
 \Omega_{1,2}$. There essentially only three qubit-oscillator states play a role in
 the dynamics: the initial no-photon state $|{\uparrow 0
 0}\rangle$ and the two one-photon states $|{\downarrow 1
 0}\rangle$ and $|{\downarrow 0 1}\rangle$. The most general normalized final state can therefore be
 written as
 \begin{eqnarray}\label{general3state}
|{\psi(\infty)}\rangle &\, = \,&
\sqrt{P_{\uparrow\to\uparrow}(\infty)}\,|{\uparrow 0
 0}\rangle  \\
 &\, +\,&\sqrt{P_{\uparrow\to\downarrow}(\infty)}\,\left(\, s_{10}|{\downarrow 1
 0}\rangle \,+ \,s_{01}|{\downarrow 0
 1}\rangle\,\right),  \nn  \end{eqnarray}
 with probabilities $P_{\uparrow\to\uparrow}(\infty)$ and $P_{\uparrow\to\downarrow}(\infty)$
 given in Eq.~(\ref{centralresulttwoosc}) and with general complex coefficients $s$ that are only constrained by
 $|s_{10}|^{2} + |s_{01}|^{2}=1$ to ensure state
 normalization. Below, we will analyze the final state in the limits of large
 and vanishing  detuning of the oscillator frequencies, before addressing
 the intermediate case $\delta\omega \simeq \gamma/\hbar$.

 \subsection{Large detuning ($\delta\omega \gg \gamma/\hbar$)}\label{sec:largedetuning}
\begin{figure}[t]
\begin{center}
\includegraphics*[width=8cm]{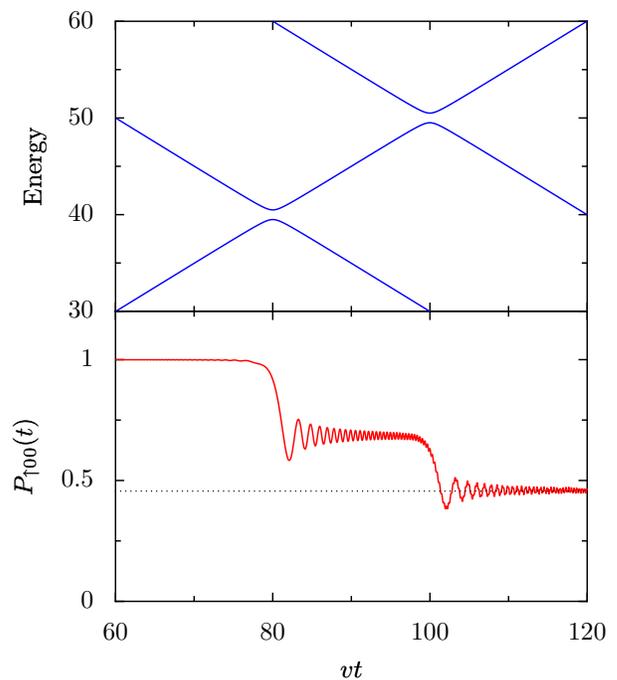} \end{center} \caption{
(Color online) Upper panel: Adiabatic energies during a LZ sweep of
a qubit
 coupled to two oscillators. Parameters:
$\gamma = 0.25\sqrt{\hbar v}$ and $\Omega_{2}= 100\sqrt{\hbar v}$,
both as in Fig.~\ref{fig:energylandscape}; $\hbar\Omega_{1}=
80\sqrt{\hbar v}$. Lower panel: Probability
$P_{\uparrow\rightarrow\uparrow}(t)$ that the system stays in the
initial state $|{{\uparrow} 0 0}\rangle$ (solid), and corresponding
exact survival final survival probability
$P_{\uparrow\rightarrow\uparrow}(\infty)$ of
Eq.~(\ref{centralresulttwoosc}) (dotted).} \label{fig:largedetuning}
\end{figure}

 If the resonance energies of the cavities differ by much more than
 the qubit-oscillator coupling, then the dynamics can very well be approximated
 by two {\em independent} standard Landau-Zener transitions, see Figure \ref{fig:largedetuning}.
 The first transition
 occurs when the qubit energy $v t$ is resonant with the energy
 $\hbar \Omega_{1}$ of the first oscillator. Only the two states $|{\uparrow 0
 0}\rangle$ and  $|{\downarrow 1  0}\rangle$ take part in this transition.
 The other  transition occurs around time $t = \hbar \Omega_{2}/v$
 and only between the two states $|{\uparrow 0
 0}\rangle$ and  $|{\downarrow 0 1}\rangle$.

This situation of two independent transitions is analogous to
optical cavity QED with two cavities and one atom flying
successively though both of them. The `time of flight' for the atom
between the two cavities here corresponds to the time difference
$\delta t = \hbar\delta\omega/v$ between the two resonances.

 Just like the
 standard LZ transition in Sec.~\ref{standard}, we can go to an interaction picture
 and summarize the transitions by S-matrices.
 In the basis
 $\{ |{\uparrow 0 0}\rangle, |{\downarrow 1 0}\rangle,|{\downarrow 0 1}\rangle \}$,
 the $S$-matrix for the first transition becomes
 \begin{equation}\label{transitionS1}
\bfsfS_{1} = \left( \begin{array}{ccc} \sqrt{q'} &
\sqrt{1-q'}\;\upe^{-\mathrm{i}\chi'-\mathrm{i}\phi_{1}} & 0
\\- \sqrt{1-q'}\;\upe^{\mathrm{i} \chi'+\mathrm{i}\phi_{1}} & \sqrt{q'} & 0 \\
0 & 0 & 1
\end{array}\right),
\end{equation}
and for the second transition
 \begin{equation}\label{transitionS2}
\bfsfS_{2} = \left( \begin{array}{ccc} \sqrt{q'} & 0 &
\sqrt{1-q'}\;\upe^{-\mathrm{i}\chi'-\mathrm{i}\phi_{2}} \\ 0 & 1 & 0
\\- \sqrt{1-q'}\;\upe^{\mathrm{i} \chi'+\mathrm{i}\phi_{2}} & 0 & \sqrt{q'}
\end{array}\right).
\end{equation}
Here the quantity $q'$ equals $\exp(-2\pi \eta')$ with adiabaticity
parameter $\eta' = \gamma^{2}/\hbar v$. The Stokes phase is now
given by $\chi' = \pi/4+\arg\Gamma(1-\mathrm{i}\eta')+\eta'(\ln
\eta'-1)$, and $\phi_{j}=\hbar\Omega_{j}^{2}/2 v$. The final state
after these two transitions is given by
$|{\tilde\psi(\infty})\rangle =
\bfsfS_{2}\bfsfS_{1}|{\tilde\psi(-\infty)}\rangle$. Hence, for the
initial state $|{\tilde\psi(-\infty)}\rangle =|{\uparrow 0
0}\rangle$ we find the final state after two transitions
 \begin{eqnarray}\label{finalstateindep}
 |{\tilde\psi(\infty)}\rangle & = & q'\,|{\uparrow 0
 0}\rangle  \\
&  -&\sqrt{1-q'} e^{\upi \theta } \left(\,|{\downarrow 1
 0}\rangle \,+ \,e^{\upi(\phi_{2}-\phi_{1})}\sqrt{q'}|{\downarrow 0
 1}\rangle\,\right), \nonumber    \end{eqnarray}
with $\theta = \chi'+\phi_{1}$. This state is a special case of
Eq.~(\ref{general3state}) in case of two independent LZ transitions.
From the final three-level state~(\ref{finalstateindep}) we find
back the {\em exact} survival probability
$|c_{\uparrow}(\infty)|^{2} = q' = \exp(-2\pi \eta)$. In the
derivation of the exact result~(\ref{centralresulttwoosc}) we took
the full Hilbert space into account rather than a three-level
subspace. Moreover,  we did not assume whether or not the LZ
transitions would occur independently. In other words, the exact
probability~(\ref{centralresulttwoosc}) comes out independent of
these two assumptions. Now that the assumptions hold, the exact
result will still hold. However, one cannot turn the argument
around: finding the survival probability $|c_{\uparrow}(\infty)|^{2}
= q' = \exp(-2\pi \eta)$ does  not imply that the rotating-wave
approximation was valid after all or that two  LZ transitions must
have occurred that were independent. Indeed, we will come across a
counterexample  in Sec.~\ref{sec:zerodetuning}.

With the final state~(\ref{finalstateindep}) determined, one can
answer the question  whether two successive and independent LZ
transitions are a suitable method for entangling the two
oscillators. Conditional on measuring the qubit in state
$|{\downarrow}\rangle$, apart from an unimportant overall phase the
entangled two-oscillator state becomes
\begin{equation}\label{conditionalstate}
|{\tilde\psi_{\rm osc}(\infty)}\rangle =
\frac{1}{\sqrt{1+q'}}\left(\,|{ 1 0}\rangle \,+
\,e^{\upi(\phi_{2}-\phi_{1})}\sqrt{q'} |{0 1}\rangle\,\right).
 \end{equation}
In general one would like to be able to create the maximally
entangled symmetric Bell state $(|{ 1 0}\rangle + |{0
1}\rangle)/\sqrt{2}$. Taking Eq.~(\ref{conditionalstate}) at face
value, the way to obtain equal probabilities for the two one-photon
states would be the case $q'=1$, but in this case the qubit would
never have ended up in $|{\downarrow}\rangle$ in the first place, as
Eq.~(\ref{finalstateindep}) reveals. So equal probabilities can not
be realized. The reason is,  for very slow LZ transitions (i.e. for
$q' \rightarrow 0$), all population already follows adiabatically
the path $|{\uparrow 0 0}\rangle \rightarrow |{\downarrow 1
0}\rangle$ in the first transition, so that no population is left to
take part in the second LZ transition and hence no population ends
up in $|{\downarrow 0 1}\rangle$. For faster transitions the
situation is less extreme, but a population difference in the final
state will remain.

In summary, entangling the two cavities via a LZ sweep of the qubit
in the case of large detuning is not ideal, since Bell states cannot
be engineered with high probability because of  a tradeoff between
the probability $P_{\uparrow\rightarrow\downarrow}(\infty)=1-q'$ to
create at least one photon and the relative probability $q'/(1+q')$
that that one photon ends up in the second cavity. On the other
hand, we find hat the state $|{\downarrow}1 0 \rangle$ can be
created with certainty in the adiabatic limit, which means that
single-photon creation in a single oscillator as discussed in
Sec.~\ref{sec:singleosc} can still be realized with LZ sweeps even
if another detuned oscillator is present; see also \cite{Lange2000a}.

\subsection{Degenerate oscillator energies ($\delta\omega \ll
\gamma/\hbar$)}\label{sec:zerodetuning}

Instead of two independent
transitions, we will now consider the other extreme case
$\delta\omega=0$ so that during the LZ sweep the qubit comes into
resonance with both oscillators at the same time. With
$\Omega_{1}=\Omega_{2}=\Omega$,  the qubit-two-oscillator system has
an extra symmetry that we will now exploit in the analysis of the LZ
dynamics. Let us first go back and not yet make the rotating-wave
approximation. The Hamiltonian~(\ref{Ht2}) now becomes
\begin{equation}\label{Ht2symm}
H = \frac{vt}{2} \sigma_{z}
   + \gamma\sigma_{x} (b_{1} + b_{1}^{\dagger}+b_{2} + b_{2}^{\dagger})
   + \hbar\Omega \left( b_{1}^{\dagger} b_{1} +  b_{2}^{\dagger} b_{2}\right).
\end{equation}
We introduce the new operators \begin{equation} b_{\pm} =
\frac{1}{\sqrt{2}} \left( \, b_{1}\pm b_{2}\,\right),
\end{equation}
which have standard commutation relations
$[b_{\pm},b_{\pm}^{\dag}]=1$ and
$[b_{\pm},b_{\pm}]=[b_{\pm},b_{\mp}^{\dag}]=0$. Both  creation
operators $b_{\pm}^{\dag}$ create one single photon with equal
probability in the first or the second oscillator:
\begin{equation}
b_{\pm}^{\dag}|{0_{+} 0_{-}}\rangle = \frac{1}{\sqrt{2}}\left(
b_{1}^{\dag} \pm b_{2}^{\dag} \right) |{0_{1} 0_{2}}\rangle =
\frac{1}{\sqrt{2}} \left(|{1_{1} 0_{2}}\rangle \pm |{0_{1}
1_{2}}\rangle \right).
\end{equation}
So $b_{+}^{\dag}$ creates the symmetric  and $b_{-}^{\dag}$ the
antisymmetric linear combination.  Instead of describing which
photon exists in which local oscillator, we can use the fact that a
general two-oscillator state can alternatively be written as
\begin{equation}
\sum_{n_{+},n_{-}=0}^{\infty} \frac{c_{n_{+}n_{-}}
(b_{+}^{\dag})^{n_{+}}(b_{-}^{\dag})^{n_{-}}|{0_{+}
0_{-}}\rangle}{\sqrt{ (n_{+})! (n_{-})! }}.
\end{equation}
 The above rewriting is useful because in terms of
the new operators, the Hamiltonian~(\ref{Ht2symm}) becomes
\begin{equation}\label{Ht2symm2nd}
H = \frac{vt}{2} \sigma_{z}
   + \sqrt{2} \gamma\sigma_{x} (b_{+} + b_{+}^{\dagger})
   + \hbar\Omega \left( b_{+}^{\dagger} b_{+} +  b_{-}^{\dagger} b_{-}\right).
\end{equation}
Note the factor $\sqrt{2}$ in the interaction term. The
Hamiltonian~(\ref{Ht2symm2nd}) shows that the qubit is fully
decoupled from the antisymmetric operators. The state of the
``antisymmetric photons'' will therefore remain unaffected by the LZ
sweep. Conversely, whatever the antisymmetric state the two
oscillators are in, it will not influence the LZ dynamics.
Consequently, for degenerate oscillator frequencies we find back the
mathematical problem for a qubit coupled to {\em one} oscillator,
which we already studied  in \cite{Saito2006a} and in Sec.
\ref{sec:singleosc} above. The difference lies in the physical
meaning of the oscillator, either a local oscillator or a symmetric
combination of two local oscillators.

\begin{figure}[t]
\begin{center}
\includegraphics*[width=8cm]{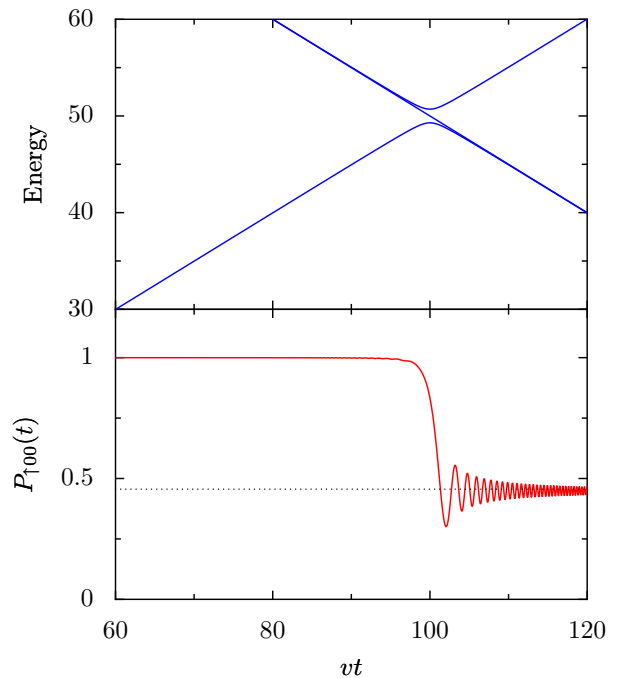} \end{center} \caption{
(Color online) Upper panel: Adiabatic energies during a LZ sweep of
a qubit
 coupled to two oscillators with degenerate energies. Parameters:
$\gamma = 0.25\sqrt{\hbar v}$ and $\hbar\Omega_{2}= 100\sqrt{\hbar
v}$, as before; this time $\hbar\Omega_{1}= \hbar\Omega_{2}$. Lower
panel: Probability $P_{\uparrow\rightarrow\uparrow}(t)$ that the
system stays in the initial state $|{{\uparrow} 0 0}\rangle$
(solid), and corresponding exact survival final survival probability
$P_{\uparrow\rightarrow\uparrow}(\infty)$ of
Eq.~(\ref{centralresulttwoosc}) (dotted).} \label{fig:zerodetuning}
\end{figure}

Now assume as before that the initial state is $|{\uparrow 0_{1}
0_{2}}\rangle =|{\uparrow 0_{+} 0_{-}}\rangle$. Moreover, we take
 $\hbar\Omega_{1,2}=\hbar\Omega \gg \gamma$ so that the rotating-wave approximation
 can be made. Then again only the
three states $|{\uparrow 0_{1} 0_{2}}\rangle$, $|{\downarrow 1_{1}
0_{2}}\rangle$, and $|{\downarrow 0_{1} 1_{2}}\rangle$ will play a
role in the LZ dynamics. Or, in our new representation, only the
{\em two } states $|{\uparrow 0_{+} 0_{-}}\rangle$ and $|{\downarrow
1_{+} 0_{-}}\rangle$. The third diabatic state $|{\downarrow 0_{+}
1_{-}}\rangle$ has a time-dependent energy $(\hbar \Omega - vt/2)$
and since it is annihilated by the RWA interaction $\sqrt{2}\gamma
(\sigma_{+}b_{+}+\sigma_{-}b_{+}^{\dag})$, within the RWA the state
$|{\downarrow 0_{+} 1_{-}}\rangle$  is an adiabatic eigenstate.
Indeed, Figure~\ref{fig:zerodetuning}
shows that far from the crossing time $v t = \hbar \Omega$ two
adiabatic energies overlap with   energy values $(\hbar \Omega - v t
/2)$, which is the energy of the diabatic states $|{\downarrow 1_{1}
0_{2}}\rangle$ and $|{\downarrow 0_{1} 1_{2}}\rangle$, or
equivalently, of $|{\downarrow 1_{+} 0_{-}}\rangle$ and
$|{\downarrow 0_{+} 1_{-}}\rangle$. At the crossing, one such a
diabatic line remains and it corresponds to the state $|{\downarrow
0_{+} 1_{-}}\rangle$. The other two states $|{\uparrow 0_{+}
0_{-}}\rangle$ and $|{\downarrow 1_{+} 0_{-}}\rangle$ form the
avoided crossing in Fig.~\ref{fig:zerodetuning}.

The LZ transition probabilities can again be calculated with the help of an
$S$-matrix. The $(2\times2)$-dimensional S-matrix in the $\{ |{\uparrow 0_{+} 0_{-}}\rangle,
|{\downarrow 1_{+} 0_{-}}\rangle \}$ basis gets the form
\begin{equation}\label{transitionS}
\bfsfS_{+} = \left( \begin{array}{cc} \sqrt{q''} &
\sqrt{1-q''}\;\upe^{-\mathrm{i}\chi''}
\\- \sqrt{1-q''}\;\upe^{\mathrm{i} \chi''} & \sqrt{q''} \end{array}\right),
\end{equation}
where $q'' = \exp(- 2 \pi \eta'')$ and the Stokes phase is given by
$\chi'' = \pi/4+\arg\Gamma(1-\mathrm{i}\eta'')+\eta''(\ln
\eta''-1)+\hbar\Omega^{2}/2 v$, with adiabaticity parameter $\eta''
= (\sqrt{2}\gamma)^{2}/\hbar v$. It immediately follows that
$|c_{\uparrow}(\infty)|^{2} = q'' = \exp(- 4 \pi \gamma^{2}/\hbar
v)$, again in agreement with the exact result of
Eq.~(\ref{centralresulttwoosc}). Thereby the same generally valid
exact result indeed shows up in the two limiting cases that can be
treated with simple S-matrices: the case ($\hbar\Omega_{1,2}\gg
\gamma, \delta\omega \gg \gamma$) in Sec.~\ref{sec:largedetuning},
and here the case  ($\hbar\Omega_{1}=\hbar\Omega_{2}\gg \gamma$).

The final three-level state in these two limiting cases is
different, though. Instead of the state~(\ref{finalstateindep}), for
the qubit coupled to degenerate oscillators starting in $|{\uparrow
0_{1} 0_{2}}\rangle$ we now find the final state
\begin{eqnarray}\label{finalstatedegenerate}
|{\tilde\psi(\infty)}\rangle & = & \sqrt{q''} |{\uparrow 0_{+} 0_{-}}\rangle -\sqrt{1 - q''} e^{i \chi''} |{\downarrow 1_{+} 0_{-}}\rangle  \\\
& = & \sqrt{q''} |{\uparrow 0_{1} 0_{2}}\rangle -\sqrt{1 - q''} e^{i \chi''}
\left( \frac{ |{\downarrow 1_{1} 0_{2}}\rangle + |{\downarrow 0_{1} 1_{2}}\rangle}{\sqrt{2}}\right). \nn
\end{eqnarray}
Clearly, if the qubit is finally measured and found in state $|{\downarrow}\rangle$, then the
two oscillators end up in the symmetric Bell state
\begin{equation}\label{Bellstate}
|{\tilde\psi_{\rm osc}(\infty)}\rangle =\frac{1}{\sqrt{2}}\left(\,
|{ 1_{1} 0_{2}}\rangle + |{ 0_{1} 1_{2}}\rangle \,\right).
\end{equation}
Notice that unlike in the final oscillator
state~(\ref{conditionalstate}) after two {\em independent} LZ
transitions, no relative phase between the two oscillator states $|{
1_{1} 0_{2}}\rangle$ and $|{ 0_{1} 1_{2}}\rangle$ is built up here.
For degenerate oscillators this is as one would expect.

It follows from Eq.~(\ref{finalstatedegenerate}) that in practice
one has two options to produce this Bell state~(\ref{Bellstate}):
The first option is to employ fast LZ transitions which produce
either the sought oscillator Bell state or the oscillator ground
state. A measurement of the qubit state is then required, as in a
proposal~\cite{Browne2003a} to create Bell states in optical
two-cavity QED. If $|{\uparrow}\rangle$ is measured, then a fast LZ
sweep back to the initial ground state is needed and the process can
be repeated until the qubit is measured in state
$|{\downarrow}\rangle$, whereby the Bell state is produced. The
second option is simpler: one chooses slow LZ transitions and by
adiabatic following produces the Bell state~(\ref{Bellstate}) with
almost certainty. The reason why the first option may be preferred
after all in practice  is that the usual fight against decoherence
may require faster than adiabatic operations.

An important difference with the final oscillator state~(\ref{conditionalstate}) after two
{\em independent} LZ transitions in Sec.~\ref{sec:largedetuning}  is that in degenerate
oscillators always the symmetric Bell state will be produced by measuring the qubit state
in $|{\downarrow}\rangle$, independent of the adiabaticity parameter
$\eta''$. So if the goal is to produce the symmetric Bell state, then one should try and
build systems where a qubit couples with equal strength $\gamma$ to two oscillators with frequency detuning $\delta\omega \ll \gamma$.
 If on the other hand one would like to be able to vary the final two-oscillator state, then detuned oscillators are to be
preferred that give two independent LZ transitions, producing the state~(\ref{conditionalstate}).

\subsection{Intermediate detuning ($\delta\omega\simeq \gamma/\hbar$)}\label{sec:intermediate}
\begin{figure}[t]
\begin{center}
\includegraphics*[width=8cm]{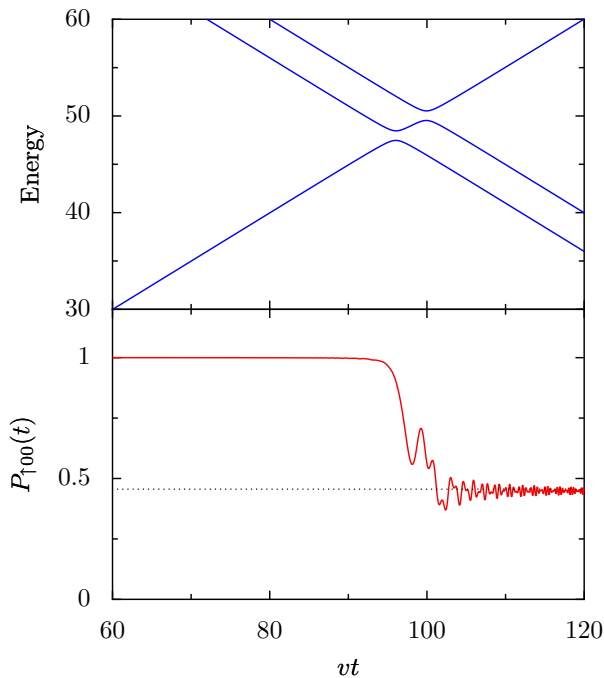} \end{center} \caption{
(Color online) Upper panel: Adiabatic energies during a LZ sweep of
a qubit
 coupled to two oscillators with large energies, and with detunings of the order of the
 qubit-oscillator coupling $\gamma$. Parameters:
$\gamma = 0.25\sqrt{\hbar v}$ and $\hbar\Omega_{2}= 100\sqrt{\hbar
v}$, as before; $\hbar\Omega_{1}= 96\sqrt{\hbar v}$. Lower panel:
Probability $P_{\uparrow\rightarrow\uparrow}(t)$ that the system
stays in the initial state $|{{\uparrow} 0 0}\rangle$ (solid), and
corresponding exact survival final survival probability
$P_{\uparrow\rightarrow\uparrow}(\infty)$ of
Eq.~(\ref{centralresulttwoosc}) (dotted). }
\label{fig:intermediatedetuning96}
\end{figure}

In case of intermediate detuning, the dynamics cannot be described
as one or as two independent successive LZ transitions (see
Fig.~\ref{fig:intermediatedetuning96}).
Of course we know from the exact result that  the survival
 $P_{\uparrow\rightarrow\uparrow}(\infty)$ probability  is given by
 Eq.~(\ref{centralresulttwoosc}), as before. Apart from this
 probability, we are again interested in the final state.
 If we assume again that $\hbar\Omega_{1,2}\gg \gamma$ and make the rotating-wave
 approximation, in the local-oscillator basis we end up with
 the $(3\times 3)$ Hamiltonian
 \begin{equation}\label{Ht_intermediate}
 H_{\rm RWA}(t)  = \left( \begin{array}{ccc}
 vt/2 & \gamma & \gamma \\
 \gamma & \hbar\Omega_{1} - vt/2 & 0 \\
 \gamma & 0 & \hbar\Omega_{2}-vt/2
 \end{array} \right).
 \end{equation}
This is a special case of the Demkov-Osherov model
\cite{Demkov1967a}, in which one level crosses $N$ parallel levels.
Interestingly,  the transition probabilities~(\ref{centralresult2})
and~(\ref{centralresulttwoosc}) are also exact within the
Demkov-Osherov model \cite{Demkov1967a,Kayanuma1985a}. Even the
scattering matrix $\bfsfS$ for the $(N+1)$ levels is known exactly
\cite{Brundobler1993a,Macek1998a}. The interesting result is that
the final state is still given by Eq.~(\ref{finalstateindep}), {\em
as if} the two level crossings had been independent. Only the phase
$\theta$ in Eq.~(\ref{finalstateindep}) is to be replaced by a more
complicated expression \cite{Macek1998a}, but for the two-oscillator
state $|{\tilde\psi_{\rm osc}(\infty)}\rangle$ this is an irrelevant
overall phase. A reason to avoid intermediate detunings in
experiments is that convergence of the relative phase to the final
value $(\phi_{2}-\phi_{1})$ in Eq.~(\ref{finalstateindep}) is
reached more slowly than for large detuning \cite{Brundobler1993a},
whereas $(\phi_{2}-\phi_{1})$ simply vanishes  for zero detuning.

In these Secs.~\ref{sec:largedetuning}-\ref{sec:intermediate} we considered
three regimes for the detuning, but we have not yet estimated
for which parameters these regimes occur. For example,
in practice the two oscillator frequencies will never be exactly
degenerate, so the question arises how to define the regime in which
they are {\em effectively} degenerate. This can be estimated by
requiring that  the typical time of a single LZ transition
$\tau_{\rm LZ}=2\gamma/v$ \cite{Mullen1989a,Wubs2005a} is much smaller than
the sweeping time $\hbar\delta\omega/v$ from energy $\hbar\Omega_{1}$ to
$\hbar\Omega_{2}$. The sweep
velocity drops out and we find the requirement $\delta\omega \ll
2\gamma/\hbar$. Following a similar reasoning, the regime of two independent LZ
transitions is characterized by $\delta\omega \gg
2\gamma/\hbar$.

\section{Experimental realisation}\label{sec:exp}
Above we  stated that the Hamiltonian~(\ref{HtNosc}) can be realized
experimentally, and here we briefly outline a realization in circuit
QED. Several realizations are promising
\cite{Blais2004a,Wallraff2004a,Chiorescu2004a,Cleland2004a,Geller2005a}, and here we
consider in more detail the setup of the experiments at Yale \cite{Blais2004a,Wallraff2004a}.

We first consider the setup with one qubit and one oscillator, which has already been realized
and for which we proposed in Ref.~\cite{Saito2006a} and in Sec. \ref{sec:singleosc}
to generate single photons with LZ sweeps.
In tunneling representation, a circuit QED setup with one
circuit oscillator is described by the Hamiltonian~\cite{Blais2004a,Wallraff2004a}
\begin{equation}
\label{ham} H(t) = -\frac{ E_\mathrm{el} }{ 2 } \sigma_{x}
    -\frac{E_\mathrm{J}(t)}{2}\sigma_{z}
    + \hbar \Omega b^{\dagger} b
    + \gamma  ( b^{\dagger} + b ) \left[\sigma_{x} +1- 2 N_\mathrm{g}\right]
\end{equation}
for the qubit, the circuit oscillator, and their mutual coupling.
The circuit cavity can indeed be modelled as a harmonic oscillator.
The electrostatic energy $E_\mathrm{el} = 4E_\mathrm{c}[1-2
N_\mathrm{g}]$ is determined by the charging energy $E_\mathrm{c}$
and the tunable gate charge $N_\mathrm{g}$.  The tunable flux
$\Phi(t)$ penetrating the superconducting loop will be used to drive
the qubit. The flux controls the Josephson energy $E_\mathrm{J}(t) =
E_\mathrm{J,max}\cos[\pi\Phi(t)/\Phi_0]$, where $\Phi_{0}$ is the
flux quantum.  The two-level approximation underlying the
Hamiltonian \eqref{ham} is valid in the ``charge regime''
$E_\mathrm{c}\gg  E_\mathrm{J}$.
In order to minimize decoherence, one typically operates the qubit
at the optimal working point $N_\mathrm{g} = \frac{1}{2}$, so that
$E_\mathrm{el}=0$ \cite{Vion2002a}. Here we also restrict ourselves
to this optimal working point.  The LZ dynamics can then be realised
by switching the flux $\Phi(t)$ in such a way that $E_\mathrm{J}(t)
= -vt$, with $v>0$. This way, our central Hamiltonian~(\ref{HtNosc}) is realized
 for one qubit coupled to one oscillator ($N=1$).
 Moreover, the very low temperatures in circuit QED experiments
\cite{Wallraff2004a} justify the assumption for our calculations that both the qubit and
the oscillator are initially in their ground states, i.e.
$|\Psi(-\infty)\rangle = |{\uparrow},0\rangle$, where $\sigma_z
|{\uparrow}\rangle = |{\uparrow}\rangle$.

 For an ideal LZ sweep, the Josephson energy should be swept from
  $E_\mathrm{J}=-\infty$ to $E_\mathrm{J}=\infty$. In practice, the two-level approximation
  is valid in a finite energy range. Moreover, in reality
  $E_\mathrm{J}$ is bounded by an $E_\mathrm{J,max}$ which
is determined by the critical current. The condition
$E_\mathrm{J,max }> \hbar \Omega$ is required so that the qubit
comes into resonance with the oscillator sometime during the sweep.
The duration of this linear sweep has to be long enough, so that
transition probabilities have converged and the finite time interval
can be extended to $t=-\infty\ldots\infty$ in calculations
describing the dynamics. In circuit QED this situation occurs, since
qubit energies can be swept around the oscillator resonance $\hbar\Omega$ over
intervals much larger than the interaction $\gamma$.

As another practical condition, inverting the flux through the
superconducting loop requires a finite time $2 T_{\mathrm{min}}$, so
that $v$ cannot exceed $v_\mathrm{max} = E_\mathrm{J,max}/2
T_{\mathrm{min}}$. For the setup of Refs.~\cite{Wallraff2004a,Blais2004a},
the sign of the initial Josephson energy $E_\mathrm{J,max} = 2\pi\hbar\times
10^{10}\mathrm{Hz}$ can be inverted within $T = 1\mu\mathrm{s}$ so
that $v_\mathrm{max} = 2\pi \hbar\times 10^{16}\mathrm{s}^{-2}$.

The cavity frequency $\Omega$ and the qubit-oscillator
coupling $\gamma$ are determined by the design of the setup.
A typical cavity frequency is $\Omega = 2\pi\times 10^9\mathrm{Hz}$.  For
the qubit-oscillator coupling strength we assume $\gamma/2 \pi \hbar
= 3\times 10^6\mathrm{Hz}$. Since $\gamma \ll \hbar \Omega$,
the generation of more than one photon is negligible. In this
limit, by choosing a proper value of $v$,  one can obtain  any
desired superposition of the states $|{\uparrow}0\rangle$ and
$|{\downarrow}1\rangle$, as explored in more detail in
\cite{Saito2006a}.

In Sec.~\ref{sec:zerodetuning} we proposed to create Bell states by coupling
 the qubit with identical couplings $\gamma$ to
two degenerate noninteracting circuit oscillators. Such a setup with
one qubit and two oscillators currently does not exist,
but a detailed proposal how one could fabricate one is given in Ref.~\cite{Storcz2006a}.
 Our wishlist of (almost) identical couplings $\gamma_{1,2}$ and
identical resonance frequencies $\Omega_{1,2}$ may not be realized
easily. The frequencies   are typically are at least $10^4$ times
larger than the couplings. Ideally one of the two oscillator
frequencies is tunable, so that it can be brought into resonance
with the other one.

\section{Summary and conclusions}\label{sec:conclusions}

We have calculated the exact Landau-Zener transition probability of
a qubit that is coupled to one or to several quantum harmonic
oscillators, given that the system starts in the ground state. The
resulting transition probability does not depend on the oscillator
frequencies.

The LZ dynamics of a qubit coupled to one oscillator can therefore
be manipulated via the sweep speed of the transition. In
state-of-the art circuit QED, both adiabatic and nonadiabatic
transitions could be realized experimentally by varying the magnetic
flux. For different speeds of the transition, the qubit and the
oscillator end up in different entangled states, so that LZ
transitions can be part of the toolbox to prepare qubit-oscillator
entangled states.

Moreover, we have shown how LZ transitions in a qubit can be
employed to entangle two oscillators. Especially for oscillators
with equal energies, it was found that LZ transitions are a robust
way to create in the oscillators the maximally entangled  state
$(|{01}\rangle + |{10}\rangle)/\sqrt{2}$  that is known as the
symmetric Bell state. The generation of this particular state
requires the qubit to undergo a LZ transition induced by both
oscillators at the same time. In circuit QED such a situation may be
engineered, for example by coupling two orthogonal transmission-line
resonators to the same superconducting qubit \cite{Storcz2006a}.

In standard cavity QED, maximally entangled
cavity states have been realized experimentally, but only for two
degenerate optical modes of the same optical cavity \cite{Rauschenbeutel2001a}.
Our proposal is different in that we consider two spatially separated circuit oscillators,
which is somewhat analogous to the optical protocol proposed in
Refs.~\cite{Browne2003a,Browne2003b}.
The difference between the present work and another recent proposal to create Bell states in
superconducting circuits \cite{Wei2006a} is that we consider entanglement creation in non-interacting cavities,
whereas in \cite{Wei2006a} it is shown how Bell states could be created in capacitively interacting qubits.

The decisive advantage of our proposal is that qubit-oscillator
interaction strengths are
 static. The archetypical way to entangle two optical cavities is by passing an
atom successively through both cavities. This leads to one
(non-)adiabatic transition followed by another. Experimentally,
there will be a spread in the velocity of the atom when repeating
the experiment. The analogue of this situation in circuit QED that
we have studied here does not suffer from this disadvantage: a
single qubit can be swept through two  oscillator resonances either
simultaneously or successively, in both cases with a well-controlled
constant sweep speed.

In principle, our proposal to couple two circuit oscillators to a
qubit to create Bell states in them can be generalized to three or
more circuit oscillators. For three frequency-degenerate oscillators
with equal coupling strength $\gamma$ to the qubit, instead of the
symmetric Bell state after the LZ sweep of the qubit, one would
obtain the so-called W state \cite{Duer2000a,Walther2005a}
\begin{equation}
|{W}\rangle = \left(|{001}\rangle + |{010}\rangle +|{100}\rangle\right)/\sqrt{3},
\end{equation}
while for $N$ oscillators the $N$-qubit W state would be produced
\cite{Haeffner2005a,Deng2006a,Doll2007a}. Needless to say that it is
technologically very challenging to fabricate such systems, but the
creation of these maximally entangled states simply by a single LZ
sweep in a qubit would be fascinating.


\section*{Acknowledgments}

We thank K. Saito and Y. Kayanuma for fruitful discussions. This
work has been supported by the DFG through SFB\,484,  SFB\,631, and
the cluster of excellence ``Nanosystems Initiative Munich''.



\end{document}